\documentclass[11pt]{article}
\usepackage[letterpaper,margin=1in]{geometry}
\usepackage[T1]{fontenc}
\usepackage[utf8]{inputenc}
\usepackage{amsmath}
\usepackage{graphicx}
\usepackage{tikz}
\usepackage{pgfplots}
\usepackage{adjustbox}
\usepackage{booktabs}
\usepackage{array}
\usepackage{tabularx}
\usepackage{fancyvrb}
\usepackage[hidelinks]{hyperref}
\hypersetup{
  pdftitle={Public-Decay Homomorphic State Space Models for Private Sequence Inference},
  pdfauthor={Luís Brito (ORCID: 0009-0006-4465-0621)},
  pdfkeywords={fully homomorphic encryption, encrypted inference, state space models, private sequence inference}
}
\pgfplotsset{compat=1.18}
\usetikzlibrary{arrows.meta,positioning,calc}
\newcolumntype{Y}{>{\raggedright\arraybackslash}X}
\newcommand{\Description}[1]{}
\title{Public-Decay Homomorphic State Space Models for Private Sequence Inference}
\author{Luís Brito\thanks{ORCID iD: \href{https://orcid.org/0009-0006-4465-0621}{0009-0006-4465-0621}}\\\small School of Technology and Management (ESTG-IPVC)\\\small Polytechnic Institute of Viana do Castelo, Viana do Castelo, Portugal}
\date{}
\begin{document}
\maketitle
\begin{abstract}

Fully homomorphic encryption (FHE) changes sequence-model design because rotations, encrypted products, ciphertext materialization, multiplicative depth, and bootstrapping pressure can dominate ordinary neural-network costs. This paper presents public-decay homomorphic state space models (HSSMs), recurrent/state-space blocks whose carried state is updated through ciphertext-plaintext public decay while ciphertext-ciphertext multiplication remains on a local write path. The design keeps a fixed encrypted state across the sequence. The evaluated workflow separates client-side tokenization, frozen fastText lookup, projection, clipping, encryption, decryption, and thresholding from server-side encrypted evaluation over bounded projected features. On full Rotten Tomatoes and SST-2 validation splits, the encrypted HSSM path exactly matches plaintext classifications and reaches 0.7505 and 0.7420 accuracy. Against HE-friendly polynomial attention on the same fastText workloads, HSSM matches or exceeds full-sequence task quality while running about $5\times$ faster. Paired L40S operation-level rows show $1.34$--$1.62\times$ lower latency than cached final-token polynomial attention, $30$--$258\times$ lower latency than full-sequence polynomial attention, and lower logical encrypted-state footprint. A $T=16/32$ comparator with encrypted public-linear input and Q/K/V projections shows projected HSSM succeeding under depth 8/ring 32768, while projected attention succeeds under depth 10/ring 65536. A matched $T=8$ OpenFHE/FIDESlib trace finishes at final level 3 and noise-scale degree 2 on both backends. These results make public-decay carry a practical FHE co-design lever for encrypted sequence inference from bounded projected features.

\end{abstract}

\noindent\textbf{ACM CCS Concepts:} Security and privacy~Privacy-preserving protocols; Computing methodologies~Neural networks.

\noindent\textbf{Keywords:} fully homomorphic encryption, encrypted inference, state space models, private sequence inference.

\section{Introduction}

Fully homomorphic encryption (FHE) offers a direct privacy promise for language inference: a client can encrypt inputs, a server can evaluate a model over ciphertexts, and only the client decrypts the output. The difficulty is that FHE changes the cost model of sequence modeling. Operations that are routine in plaintext neural networks, such as non-polynomial activations, softmax, layer normalization, matrix multiplication, rotations, packing conversions, and context-wide attention, become first-order systems constraints under encrypted execution. Recent HE/FHE language-model systems such as THE-X, Powerformer, EncryptedLLM, MOAI, Cachemir, AEGIS, and PriTran are predominantly transformer-shaped responses to those constraints \cite{chen2022thex,park2025powerformer,decastro2025encryptedllm,zhang2026moai,yu2026cachemir,gong2026aegis,mu2026pritran}.

This paper studies a focused architectural question: can a recurrent or state-space-style block be designed so that its encrypted carry path avoids ciphertext-ciphertext multiplication through time? A naive encrypted selective recurrence can still be expensive under FHE, because an encrypted input-dependent carry gate multiplied by an encrypted carried state consumes multiplicative depth at each step. HSSM instead fixes the recurrent carry to a public/plaintext decay and confines required ciphertext-ciphertext multiplication to a local write path:

\[
h_t = a\,h_{t-1} + g(x_t)u(x_t)
\]

Here $h_t$ is the encrypted recurrent state, $a$ is public/plaintext, $g(x_t)$ and $u(x_t)$ are encrypted low-degree gate and write values, and client-side output handling remains outside the server-side homomorphic computation. In the measured task workflow, this boundary is concrete: tokenization, frozen fastText lookup, train-normalized random projection, clipping, encryption, decryption, and thresholding are client-side; the server receives encrypted bounded projected features. For the existing augmented-HSSM L40S task rows, each example enters the encrypted runner as $T=4$ chunks of width 128, or 512 bounded projected scalars, and server encrypted HSSM evaluation accounts for 10.61--10.65\% of measured encryption/evaluation/decryption component time. The resulting design contribution is a public-decay recurrent/state-space-style block for FHE-oriented server-side sequence evaluation from this bounded-feature interface, with a clear encrypted carry-path depth argument when the public-decay invariant is preserved.

The contribution is fourfold. First, the paper specifies the minimal HSSM block and its interface assumptions, including public model parameters, encrypted state, client-side preprocessing, and client-side output handling. Second, it compares HSSM symbolically with transformer-style encrypted inference and a naive encrypted selective recurrence in terms of state/cache growth, ciphertext-ciphertext products, rotations, and multiplicative-depth pressure. Third, it reports real FHE artifacts across OpenFHE CPU and FIDESlib CUDA, including L40S measurements and a matched OpenFHE/FIDESlib level trace for the same expanded HSSM path. Fourth, it separates task quality from systems cost: audited toy-feature runs are treated as latency/null-quality telemetry, the repaired fastText branch supplies the main encrypted task evidence, and the attention comparators now include both polynomial readouts and an operation-level projected-Q/K/V attention block.

The strongest task-level signal comes from the repaired L40S fastText workflow. On Rotten Tomatoes and GLUE/SST-2, the selected HSSM encrypted forward path exactly matches plaintext classifications and reaches 0.7505 and 0.7420 accuracy, respectively. A joint fastText comparison then evaluates HSSM against two HE-friendly polynomial-attention readouts under the same bounded-feature task boundary. In this workflow, HSSM beats full-sequence polynomial attention on both datasets, uses a smaller logical encrypted-state footprint, and is about $4.9\times$ faster in encrypted evaluation. Cached final-token polynomial attention remains a strong short-context readout, especially on Rotten Tomatoes, which makes the result more informative: the public-decay design advantage appears when the comparator must represent sequence structure rather than only a final-token summary.

The operation-level evidence addresses the stronger comparator question at block scope. The projected-attention artifact includes homomorphic public-linear input and Q/K/V projections from encrypted bounded features at $T \in \{16,32\}$. Under the measured L40S profiles, corrected projected HSSM succeeds at depth 8/ring 32768, while the encrypted-Q/K/V attention block succeeds at depth 10/ring 65536; in those rows, HSSM is faster and uses lower peak process GPU memory. A matched T=8 expanded HSSM trace also reconciles OpenFHE CPU and FIDESlib CUDA metadata: both backends finish the same path at final level 3 and noise-scale degree 2, while the earlier OpenFHE zero-remaining-level observation belongs to a different streaming recurrent schedule. Taken together, these results establish HSSM as a concrete FHE co-design pattern for server-side encrypted sequence computation from bounded features. The paper reserves causal-language-modeling extensions for a separate fixed-protocol encrypted LM study, keeping the present contribution centered on the measured sequence-inference path and its attention comparators.

\section{Background and Related Work}

The recent private language-inference literature under HE/FHE is largely organized around making transformer blocks executable under encryption. THE-X frames encrypted transformer inference around polynomial approximations for operations that are not directly supported by standard HE tooling, including GELU, softmax, and layer normalization \cite{chen2022thex}. Powerformer continues that adaptation line by replacing softmax and layer normalization with HE-friendlier functions and optimizing homomorphic matrix multiplication \cite{park2025powerformer}. EncryptedLLM studies GPU-accelerated FHE execution for an encrypted GPT-2 forward pass and activation approximations \cite{decastro2025encryptedllm}.

Later systems sharpen the bottleneck picture. MOAI targets packing flow, rotation-free softmax and layer-normalization algorithms, and reduced rotations in transformer-module execution \cite{zhang2026moai}. Cachemir focuses on autoregressive decoding and the transformer KV-cache interaction with HE packing and bootstrapping placement \cite{yu2026cachemir}. AEGIS studies long-sequence encrypted transformers across multiple GPUs, where encoded weights and encrypted activations stress single-GPU memory and communication \cite{gong2026aegis}. PriTran reports CPU-side CKKS transformer acceleration by reducing rotations and multiplications in ciphertext-plaintext and ciphertext-ciphertext matrix operations \cite{mu2026pritran}. Together these works support the motivation for HSSM: transformer-shaped encrypted inference is possible, but the cost model is dominated by encrypted approximation, rotations, packing/layout, matrix operations, cache behavior, and long-context memory. For these 2026 references, Cachemir is cited as an under-review arXiv preprint, AEGIS as an arXiv paper with an accepted-at-ICS-2026 source note, and PriTran with DOI-backed CGO 2026 proceedings metadata.

The novelty boundary for HSSM is the public-decay carry mechanism, considered against both prior encrypted recurrence and transformer-shaped inference. The local novelty-boundary evidence registers encrypted GRU, LSTM, and RNN/control work, including CRYPTOGRU, PrivLSTM, and HE-RNN/control. It also identifies inhibitor transformers and gated RNNs as a close architectural precedent: that work redesigns sequence models around FHE primitive costs and explicitly blocks a broad first FHE-native sequence architecture claim \cite{brannvall2023inhibitor}.

Plaintext state-space models and Mamba-style selective SSMs provide motivation, not direct FHE evidence. S4, DSS, S5, Mamba, and Mamba-2 show why fixed-state or recurrent sequence processing can be attractive in plaintext settings \cite{gu2022s4,gupta2022dss,smith2023s5,gu2023mamba,dao2024mamba2}. However, plaintext Mamba selectivity does not automatically transfer to FHE: if input-dependent encrypted gates multiply encrypted carried state, the carry path can consume multiplicative depth through time. MPCMamba is also relevant as privacy-preserving Mamba context, but it is an SMPC system rather than an FHE result \cite{liu2025mpcmamba}. HSSM is therefore best understood as state-space-inspired FHE co-design rather than "Mamba under FHE."

This related-work boundary shapes the paper's claims. We use transformer FHE systems to motivate the encrypted attention and cache problem, encrypted recurrent systems to avoid overclaiming novelty, inhibitor-style sequence redesign to delimit FHE-native architecture claims, and plaintext SSM/Mamba work to motivate fixed-state sequence modeling. This positioning yields the paper's central empirical question: whether a public-decay recurrent carry gives useful systems behavior under registered FHE artifacts.

\section{Problem Formulation}

We consider a single-client semi-honest encrypted-inference setting. The client performs plaintext preprocessing before encryption and holds the secret key, decrypted outputs, and all decoding or thresholding logic. In the current task evidence, this boundary includes tokenization, fastText lookup, projection, and clipping; encrypted server-side computation begins only after those steps and encryption, with bounded projected features as ciphertext inputs. The server holds public/plaintext model parameters, evaluation keys, encrypted inputs, encrypted recurrent state, encrypted intermediates, and encrypted outputs before return to the client. The server is assumed to follow the specified homomorphic computation, but it may observe operation shapes, ciphertext counts, timing, memory use, failures, sequence length, and other metadata. Malicious-server security, side-channel resistance, model privacy, homomorphic tokenization or embedding lookup, fully homomorphic decoding, full raw-token-to-logit private serving, and production-serving claims are outside the current evidence.

The relevant cost units are not only arithmetic counts. For FHE sequence models, the design must track ciphertext-ciphertext multiplications, ciphertext-plaintext operations, rotations and slot permutations, packing and layout transitions, multiplicative depth or remaining levels, bootstrapping or refresh requirements, encrypted state/cache size, key material, and hardware memory pressure. The symbolic evidence uses $T$ for context length, $L$ for layers or blocks, $H$ for heads, and ciphertext-count functions that depend on packing layout. Exact constants are backend-specific. Client preprocessing and client cryptography are separate workload components, not part of the server-side encrypted-evaluation cost term.

Transformer-style encrypted inference has two distinct pressures. First, the attention and feed-forward computations require encrypted products, public linear maps, rotations, and approximations for operations such as softmax, normalization, and activations. Second, cached autoregressive decoding carries context-growing material. A symbolic KV-cache footprint has the form $L H T (C_K + C_V)$ up to layout and packing details, and a fully materialized attention-score matrix has $T^2$ score units. This paper does not claim that transformer depth must grow linearly with $T$; rather, it treats work, cache footprint, rotations, packing pressure, and memory as context-dependent terms.

Naive encrypted selective recurrence avoids transformer attention, but it can place a ciphertext-ciphertext product on the recurrent carry path. A typical form is:

\[
h_t = g_A(x_t)\,h_{t-1} + g_B(x_t)\,u(x_t)
\]

When $g_A(x_t)$ and $h_{t-1}$ are both encrypted, the carry term consumes an encrypted multiplication at every recurrent step. The registered formal analysis summarizes this pressure as:

\[
d_h(t) = \max\{\max(d_g, d_h(t-1)) + 1, d_w\}
\]

where the $+1$ comes from multiplying the encrypted carry gate into the encrypted carried state. Bootstrapping or refresh can bound this growth over an interval, but then the cost shifts to the refresh schedule and backend support.

The HSSM problem is to preserve an encrypted recurrent state while removing ciphertext-ciphertext multiplication from the recurrent carry path. The minimal update is:

\[
h_t = a\,h_{t-1} + g_t u_t
\]

where $a$ is public/plaintext. This does not eliminate encrypted multiplication: the local write $g_t u_t$ remains encrypted. It does, however, change the recurrent dependency from encrypted gate times encrypted state to plaintext decay times encrypted state. The paper asks whether this structural change is visible in symbolic analysis, real FHE loops, GPU kernels, and bounded end-to-end workflows.

**Assumption 1 (public-decay carry invariant).** The recurrent carry path is eligible for the HSSM carry-depth argument only when the coefficient multiplying the carried state is public/plaintext at inference. Under this invariant, $a h_{t-1}$ is a ciphertext-plaintext operation. If a variant replaces $a$ with an encrypted input-dependent carry gate multiplied into $h_{t-1}$, that variant falls back to the naive encrypted carry-path analysis above.

\section{Public-Decay HSSM Architecture}

The server-side interface begins after feature construction. In the current fastText/L40S task artifact, the client performs raw-text tokenization, fastText lookup, train-normalized random projection, clipping, and encryption before the server sees an input. The server receives encrypted bounded projected feature chunks, not raw text, token IDs, or fastText vectors. For the existing augmented-HSSM task rows this payload is $T=4$ chunks of width 128, or 512 bounded projected scalars per example. Thus $\tilde{x}_t=\operatorname{Enc}(x_t)$ below denotes an encrypted bounded projected feature chunk. This section does not claim homomorphic tokenization, homomorphic fastText lookup, encrypted raw-token-to-logit inference, model-private weights, or fully homomorphic output decoding.

The minimal HSSM block operates over these encrypted bounded feature inputs and encrypted recurrent state $\tilde{h}_t = \operatorname{Enc}(h_t)$. Model parameters are public/plaintext in the current design. This includes any server-side affine feature map, the gate and write projections, recurrent decay, and output projection. Encrypted model weights and model privacy are outside the current artifact.

For each step, the server first evaluates a public affine projection:

\[
\tilde{z}_t = A_x \tilde{x}_t + b_x
\]

It then evaluates low-degree encrypted gate and write paths:

\[
\begin{aligned}
\tilde{r}_t = G \tilde{z}_t + b_g
\\
\tilde{g}_t = \phi_g(\tilde{r}_t)
\\
\tilde{s}_t = U \tilde{z}_t + b_u
\\
\tilde{u}_t = \phi_u(\tilde{s}_t)
\\
\tilde{w}_t = \tilde{g}_t \tilde{u}_t
\end{aligned}
\]

The recurrent carry coefficient $a$ is public/plaintext. The state update is:

\[
\tilde{h}_t = a\,\tilde{h}_{t-1} + \tilde{w}_t
\]

The output path is a public affine map:

\[
\tilde{y}_t = C \tilde{h}_t + b_y
\]

The block diagram summarizes the server-side update from encrypted bounded projected features.

\begin{figure}[ht]
\centering
\includegraphics[width=0.95\linewidth]{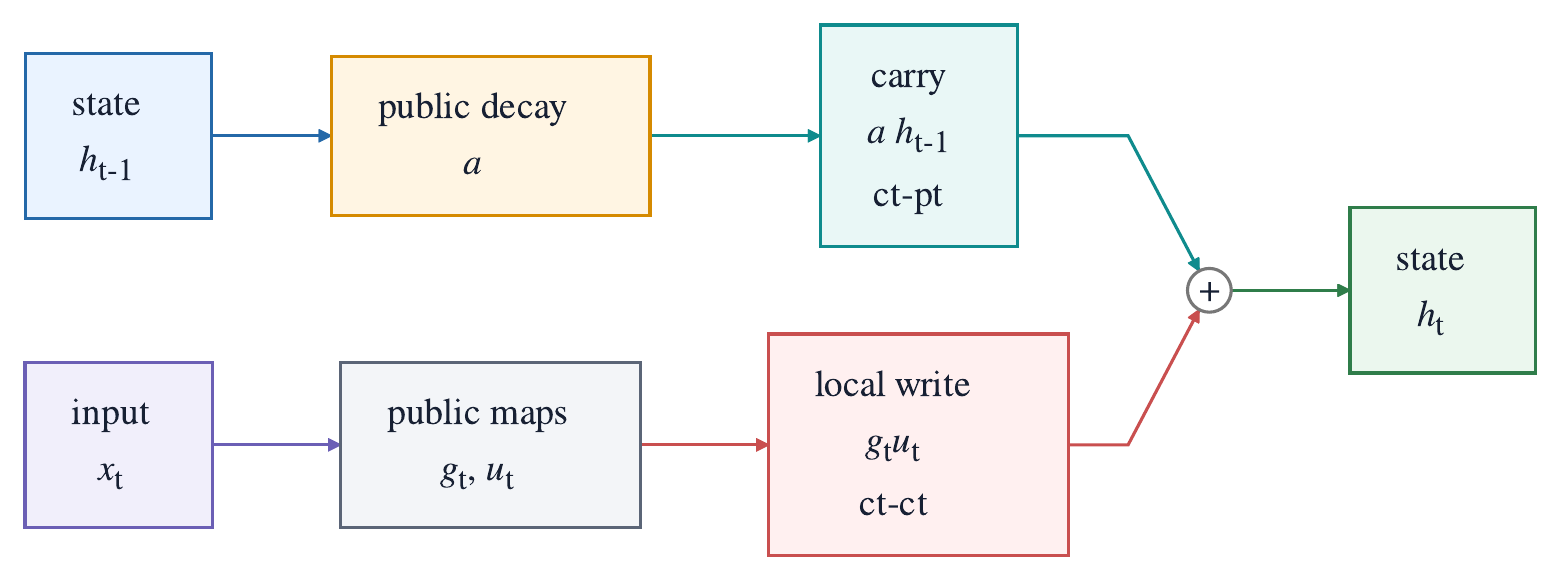}
\caption{Server-side HSSM update from encrypted bounded projected features. Public/plaintext decay carries encrypted state; the only ciphertext-ciphertext product is the local gate/write update.}
\Description{Server-side HSSM update from encrypted bounded projected features. Public/plaintext decay carries encrypted state; the only ciphertext-ciphertext product is the local gate/write update.}
\end{figure}

The server returns encrypted output material to the client. Decryption, argmax, softmax, sampling, top-k, temperature scaling, detokenization, or task-specific thresholding are client-side operations in the supported design. For the existing augmented-HSSM L40S rows, the client/server workload accounting places server encrypted HSSM evaluation at 10.61--10.65\% of the measured encryption/evaluation/decryption component time. This is workload accounting for the partition only; it does not add a stronger privacy claim.

The central invariant is therefore not an incidental implementation detail. If a future variant replaces the public decay with an encrypted input-dependent carry gate multiplied by the encrypted state, it no longer inherits the bounded encrypted carry-path depth argument. The architecture evidence treats public decay, low-degree bounded gate/write maps, and state-norm control as design conditions.

The multi-decay extension used in later controlled experiments keeps the same invariant but carries several public-decay state tracks:

\[
h_t^{(k)} = a_k h_{t-1}^{(k)} + g_t u_t
\]

A public readout combines the state bank after the sequence. This tests the expressivity-latency tradeoff of the public-decay design space without introducing encrypted carry-gate times encrypted state. The cost is direct: a bank with more decays uses more encrypted state tracks and more encrypted evaluation work. The experiments therefore treat multi-decay as a diagnostic ablation, not as established evidence of accuracy superiority.

\section{Complexity and Noise Analysis}

The symbolic comparison separates three designs: HE-friendly transformer inference, naive encrypted selective recurrence, and public-decay HSSM. The transformer path carries context-dependent work and cache material. The naive recurrent path can keep a fixed-size state but may consume multiplicative depth through time if an encrypted carry gate multiplies the encrypted state. The HSSM path keeps a fixed-size encrypted state and makes the recurrent carry a ciphertext-plaintext operation when Assumption 1 is preserved.

For transformer-style inference, the key symbolic pressures are encrypted attention products, polynomial replacements for non-polynomial functions, rotations, packing conversions, public and encrypted matrix operations, bootstrapping placement, and context-growing cache material. For the naive selective recurrence, the symbolic pressure is concentrated on the carry: $g_A(x_t)h_{t-1}$ consumes a fresh encrypted multiplication whenever both factors are encrypted. For HSSM, the local write $g_t u_t$ still consumes an encrypted multiplication, but the recurrent carry $a h_{t-1}$ is ciphertext-plaintext.

\subsection{Carry-path multiplicative depth versus CKKS level consumption}

The first claim is algebraic. Under Assumption 1, HSSM removes ciphertext-ciphertext multiplication from the recurrent carry path. The carry term $a h_{t-1}$ is ciphertext-plaintext by construction, while the required ciphertext-ciphertext product is the local write $g_t u_t$ at the current step.

The second observation is backend-specific. The older OpenFHE CPU recurrent-loop evidence does not show measured constant-library-level recurrence: under the tested OpenFHE 1.5.1 CKKS calls, reported level metadata advanced across HSSM steps and reached zero estimated levels remaining at $T=8$. That row remains useful diagnostic evidence, but it is a different streaming gate/write-vector schedule from the expanded-polynomial HSSM path used in the newer cross-backend trace.

The matched trace isolates the circuit question cleanly. With the same deterministic bounded inputs, exact expanded local polynomial, balanced public-decay aggregation, depth 8, scale 50, ring dimension 32768, and $T=8$, OpenFHE CPU and FIDESlib CUDA/L40S finish the expanded-HSSM path at the same final level and noise-scale degree, as summarized in Table~\ref{tab:unified-level-trace}.

\begin{table*}[t]
\centering
\scriptsize
\caption{Unified T=8 level trace for the matched expanded-HSSM path. The row is level-metadata and numerical-fidelity evidence only; the two runners use different layouts.}
\label{tab:unified-level-trace}
\setlength{\tabcolsep}{3pt}
\renewcommand{\arraystretch}{1.15}
\begin{tabularx}{\textwidth}{YYYY}
\toprule
Backend & Layout & Final level & Noise-scale degree \\
\midrule
OpenFHE CPU & packed 8-slot ciphertext & 3 & 2 \\

FIDESlib CUDA/L40S & eight scalar ciphertexts & 3 & 2 \\
\bottomrule
\end{tabularx}
\end{table*}

This row is level-metadata and numerical-fidelity evidence for the matched circuit. Throughput and bootstrapping schedules are measured by the separate runtime and depth-budget artifacts because the two runners use different layouts.

The third observation is measured systems behavior. In the same OpenFHE recurrent-loop artifact, the $T=8$ HSSM public-scalar carry operations sum to 23.710202 ms, while the naive encrypted carry-gate-times-state operations sum to 165.264703 ms. Thus the current evidence supports reduced encrypted carry-product pressure and a measured carry-operation latency advantage, while leaving concrete CKKS noise advantages and constant-library-level backend behavior unproven.

The repaired fastText L40S path adds a bounded-feature depth and bootstrapping budget for the supported HSSM server-side subgraph. After client-side tokenization, fastText lookup, random projection, and clipping, the encrypted path covers bounded feature encryption, low-degree gate/write maps, the local $g_tu_t$ write, public-decay aggregation, and public readout. The static trace places the shared square and gate/write polynomials at depth 1, the local write at ciphertext-ciphertext depth 2, and the public-decay state aggregation at the same ciphertext-ciphertext depth because the decay factors are public. The registered fastText validation rows complete under the depth-8 FIDESlib profile without invoking or measuring CKKS bootstrapping, and the final score telemetry reports level 4 and noise-scale degree 2 for all rows. This gives a clear no-bootstrap result for the measured bounded-feature HSSM path; measured bootstrap schedules, bootstrap latency, and raw-token-to-logit depth remain separate scale-up questions.

The encrypted-Q/K/V comparator evidence sharpens the attention side of this budget. In the L40S operation-level artifact, the corrected projected-HSSM block succeeds at $T=16$ and $T=32$ under depth 8 and ring dimension 32768. The final-token polynomial-attention block includes homomorphic public-linear input and Q/K/V projections from encrypted bounded features, and succeeds at the same two sequence lengths under depth 10 and ring dimension 65536. This supports a profile-specific projection-cost comparison while keeping the comparator class explicit: polynomial attention/readout computation rather than exact softmax, feed-forward layers, residual paths, layer normalization, model-private weights, task accuracy, or full transformer-block behavior.

The memory distinction is sharper at the model level. At fixed experimental width, the streaming HSSM state is one ciphertext independent of sequence length $T$. The implemented final-token attention surrogate retains $T$ encrypted feature-cache ciphertexts; a standard K/V cache would retain at least $2T$; and a fully materialized attention-score matrix uses $T^2$ ciphertext units. This $T^2$ score footprint is a memory/materialization axis, not a statement that transformer multiplicative depth itself grows as $T^2$. In the registered memory-scaling artifact for the depth-8, ring-32768 profile, $T=1066$ corresponds to one HSSM state ciphertext, 1066 implemented surrogate feature-cache ciphertexts, 2132 standard K/V-cache ciphertexts, and 1136356 full attention-score ciphertext units. These are model-level footprint estimates, not isolated GPU allocator measurements.

The memory-scaling figure below visualizes this distinction. It is an accounting plot over ciphertext units, not a replacement for the measured GPU stress sweep. Its leftmost tick, $T=3$, reflects the current full-validation executable's fixed sequence length; larger $T$ values in the plot are model-level context-scaling and stress-sweep accounting rather than additional full-validation sequence lengths.

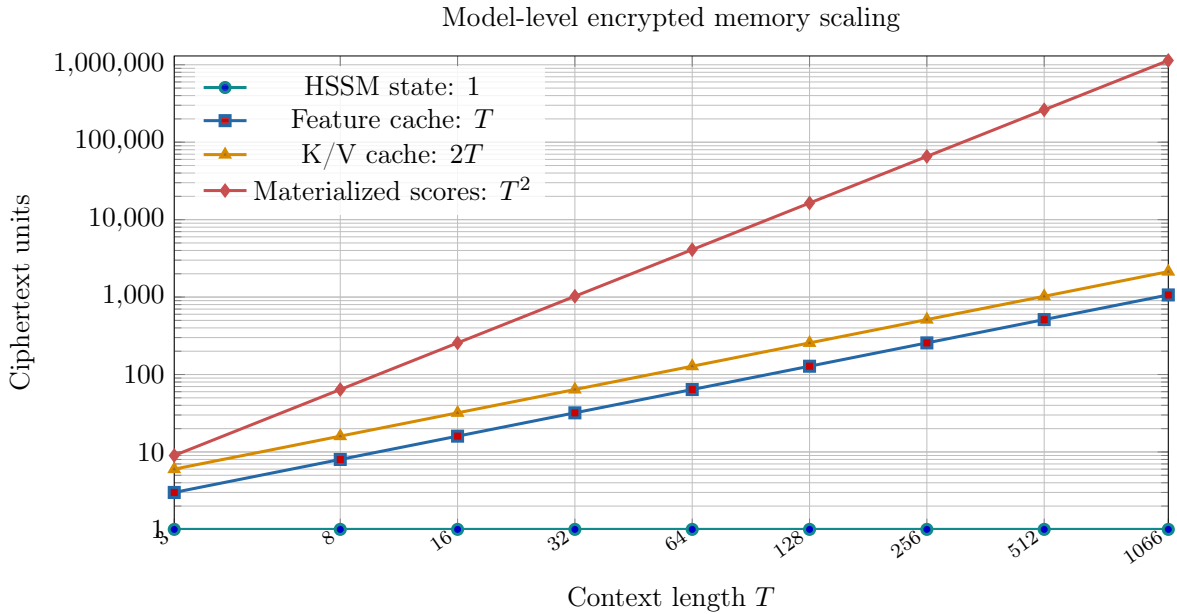
\begin{figure}[ht]
\centering
\begin{adjustbox}{max width=0.95\linewidth}
\begin{tikzpicture}
\definecolor{hssmteal}{HTML}{0F8B8D}
\definecolor{hssmblue}{HTML}{2468A8}
\definecolor{hssmamber}{HTML}{D58A00}
\definecolor{hssmred}{HTML}{C94F4F}
\begin{loglogaxis}[
  width=15.5cm,
  height=8.2cm,
  xlabel={Context length $T$},
  ylabel={Ciphertext units},
  xmin=3, xmax=1066,
  ymin=1, ymax=1300000,
  xtick={3,8,16,32,64,128,256,512,1066},
  xticklabels={3,8,16,32,64,128,256,512,1066},
  xticklabel style={rotate=35, anchor=east, font=\scriptsize},
  ytick={1,10,100,1000,10000,100000,1000000},
  log ticks with fixed point,
  grid=both,
  legend style={at={(0.02,0.98)}, anchor=north west, draw=none, fill=white, fill opacity=0.85, text opacity=1},
  title={Model-level encrypted memory scaling},
]
\addplot+[hssmteal, very thick, mark=*] coordinates {
  (3,1) (8,1) (16,1) (32,1) (64,1) (128,1) (256,1) (512,1) (1066,1)
};
\addlegendentry{HSSM state: $1$}
\addplot+[hssmblue, very thick, mark=square*] coordinates {
  (3,3) (8,8) (16,16) (32,32) (64,64) (128,128) (256,256) (512,512) (1066,1066)
};
\addlegendentry{Feature cache: $T$}
\addplot+[hssmamber, very thick, mark=triangle*] coordinates {
  (3,6) (8,16) (16,32) (32,64) (64,128) (128,256) (256,512) (512,1024) (1066,2132)
};
\addlegendentry{K/V cache: $2T$}
\addplot+[hssmred, very thick, mark=diamond*] coordinates {
  (3,9) (8,64) (16,256) (32,1024) (64,4096) (128,16384) (256,65536) (512,262144) (1066,1136356)
};
\addlegendentry{Materialized scores: $T^2$}
\end{loglogaxis}
\end{tikzpicture}
\end{adjustbox}
\caption{Model-level encrypted memory scaling. HSSM retains one recurrent state ciphertext, while transformer cache and materialized attention score footprints grow with context length.}
\Description{Model-level encrypted memory scaling. HSSM retains one recurrent state ciphertext, while transformer cache and materialized attention score footprints grow with context length.}
\end{figure}

The GPU stress evidence gives an observed counterpart to that model-level argument. Under the fixed depth-8, scale-50, batch-8, ring-dimension-32768 FIDESlib CUDA profile on the RTX 4060, materialized quadratic attention completes through $T=32$ with 1024 encrypted score ciphertexts and 7379 MiB peak polled GPU memory, then fails at $T=36$ while attempting 1296 score ciphertexts. The L40S rerun keeps the same prepared stress workload and CKKS profile, completes materialized quadratic attention at $T=36$, and records 9783 MiB peak monitored GPU memory for that row. Physical peak-memory claims are not drawn from task-table peak columns, the model-level ciphertext accounting figure, or paired operation-level polynomial-attention rows. In the paired operation rows, FIDESlib allocator pooling keeps process peaks close, so those rows support latency and logical encrypted-state claims only. Physical peak-memory superiority is claimed only for the materialized stress rows where the measured HSSM and quadratic-attention footprints separate; the near-equal $T=4$ and $T=8$ rows are allocator-floor behavior rather than meaningful separation. Adaptive-depth naive probes recover rows through $T=24$, but only by changing depth and ring dimension, and $T=32$ depth-36 fails with CUDA out-of-memory.

The FHE-aware simulation adds a numerical proxy rather than real-noise proof. Under the nominal 50-bit proxy through $T=128$, both HSSM and the naive analogue record zero clip events and zero instability-proxy counts. Under the tight state-clipping stress, HSSM records fewer clip events than the naive analogue but is not clip-free. This supports a bounded dynamic-range discussion, not real CKKS noise behavior, concrete bootstrapping, or constant-library-level backend claims.

\section{Implementation}

The implementation evidence is intentionally layered. A plaintext structural artifact defines three model paths: HSSM, a naive selective recurrent analogue, and a softmax-free final-token attention surrogate. A CKKS-style simulation tests rounding, clipping, and proxy stability. OpenFHE CPU artifacts provide real FHE single-cell and recurrent-loop checks. FIDESlib CUDA artifacts provide GPU execution for recurrent updates, cached and projected polynomial-attention comparators, server-side encrypted forward-path workflows from bounded projected features, and stress sweeps.

The local OpenFHE parameter/keygen/smoke artifact uses OpenFHE v1.5.1 CKKS and records that secret key material was generated only in process memory; no serialized secret key material was written to the repository. The OpenFHE single-cell benchmark evaluates the packed operation:

\[
h_{\mathrm{next}} = a\,h + g\,u
\]

on packed 8-slot encrypted vectors. The OpenFHE recurrent-loop artifact then compares HSSM public decay and naive encrypted selective recurrence over $T \in \{1,2,4,8\}$ on CPU under a depth-8 profile.

The first GPU path uses a cache-local FIDESlib CUDA setup on the local RTX 4060. The environment evidence records Ubuntu 24.04.4, an Intel i9-12900H CPU, 31 GiB RAM, an NVIDIA GeForce RTX 4060 with 8188 MiB reported VRAM, NVIDIA driver 580.126.09, and CUDA Toolkit 13.2 through \texttt{/usr/local/cuda/bin/nvcc}. The backend is viable but fragile: the earlier CUDA 12.0 build did not compile the full FIDESlib library, while the CUDA 13.2 retry required explicit CUDA pathing, the CUDA 13 CCCL include path, patched OpenFHE 1.4.2, and a cache-local CUDA graph/capture compatibility shim. The bundled FIDESlib test target remained blocked by missing NCCL headers.

A later scale-up evidence round staged the same Ada-targeted FIDESlib runner binaries and CUDA runtime on a remote NVIDIA L40S host. That host reported 46068 MiB VRAM and the same 580.126.09 driver family. This was a binary rerun, not a fresh remote FIDESlib build, so it supports hardware-headroom comparison for the existing proof-of-concept workloads rather than a new backend portability claim.

Table~\ref{tab:implementation-artifacts} consolidates the encrypted/backend artifacts used by the current manuscript claims. The plaintext structural model defines the HSSM, naive recurrent analogue, and attention-surrogate shapes with degree-2 polynomials; the table focuses on the checks that carry encrypted execution, backend, and profile claims. Rows marked as operation-level should not be read as full encrypted language-model executions.

\begin{table*}[t]
\centering
\scriptsize
\caption{Compressed implementation artifact map, backend profiles, supported roles, and key recorded checks used by the manuscript claims.}
\label{tab:implementation-artifacts}
\setlength{\tabcolsep}{3pt}
\renewcommand{\arraystretch}{1.15}
\begin{tabularx}{\textwidth}{>{\hsize=0.55\hsize\linewidth=\hsize\raggedright\arraybackslash}X>{\hsize=0.65\hsize\linewidth=\hsize\raggedright\arraybackslash}X>{\hsize=1.30\hsize\linewidth=\hsize\raggedright\arraybackslash}X>{\hsize=1.50\hsize\linewidth=\hsize\raggedright\arraybackslash}X}
\toprule
Artifact cluster & Backend/profile & Supported role & Key checks and scope \\
\midrule
OpenFHE CPU references & OpenFHE 1.5.1 CKKS, \texttt{HEStd\_128\_classic}, \texttt{FLEXIBLEAUTO} & Checks parameter/key generation, packed single-cell execution, and the HSSM-vs-naive recurrent loop. & Smoke: depth 2, scale 50 bits, batch 8, ring 16384, about $5\times10^{-12}$ max error. Loop: depth 8, scale 50 bits, batch 8, ring 32768 over $T\in\{1,2,4,8\}$, at most about $9\times10^{-12}$ final-state error. \\

FIDESlib recurrent update & FIDESlib 2.1.0 CUDA 13.2 with patched OpenFHE 1.4.2 on RTX 4060 & Measures operation-level HSSM and naive recurrent updates with plaintext-generated gate/write vectors encrypted as step inputs. & Depth 8, scale 50 bits, batch 8, ring dimension 32768; one discardable HSSM warm-up row per horizon; max decrypt error at most $3.23\times10^{-13}$; 43 reported precision bits. \\

Attention/projection comparators & FIDESlib CUDA L40S depth-8 and depth-10 profiles & Measures cached polynomial attention, encrypted public-linear Q/K/V projection cost, and the corrected projected-HSSM block. & Cached rows use $1+z+0.5z^2$ and $1-r+r^2$. Projected attention succeeds at $T=16,32$ under depth 10/ring 65536; corrected projected HSSM succeeds at $T=16,32$ under depth 8/ring 32768. See Table~\ref{tab:qkv-projection-comparator}. \\

Repaired fastText HSSM path & FIDESlib CUDA L40S, selected depth-8 profiles & Runs encrypted HSSM validation from bounded projected fastText features through public readout. & Client-side tokenization, fastText lookup, projection, clipping, decryption, and thresholding; encrypted server-side bounded features, gate/write polynomials, local writes, public-decay aggregation, and public readout; $T=4$, width 128, and 512 bounded projected scalars per example enter the encrypted runner. \\

Validation, stress, and L40S reruns & FIDESlib CUDA depth-8, scale-50, batch-8 family on RTX 4060 and L40S & Runs audited toy-feature telemetry, repaired fastText validation, multi-decay decision rows, materialized quadratic-attention stress, selected repeats, and scale-up reruns. & Old toy-feature accuracy demoted; fastText encrypted classifications exactly match plaintext; stress sweep stores explicit $T^2$ score ciphertexts; L40S completes HSSM and materialized quadratic attention through $T=36$; paired rows support latency and logical-state comparisons. \\
\bottomrule
\end{tabularx}
\end{table*}

This compressed map is intentionally conservative about level accounting. The matched level trace is reported separately in Table~\ref{tab:unified-level-trace}: once the circuit and schedule are identical, OpenFHE and FIDESlib run the expanded local polynomial and balanced public-decay aggregation to the same final level at $T=8$. The repaired fastText HSSM path is stronger than the recurrent-update-only artifact: after the client-side feature boundary, it evaluates encrypted bounded features, gate/write polynomials, local writes, public-decay aggregation, and a homomorphic public readout. Its depth audit shows that the supported path fits inside the depth-8 profile without measured bootstrapping. It still does not implement homomorphic tokenization, homomorphic embedding lookup, model-private learned projection from raw embeddings, model-private weights, or homomorphic output decoding.

The FIDESlib HSSM recurrent-update benchmark is recurrent-update-only. It uses plaintext-generated gate/write vectors encrypted as step inputs; it is not a full encrypted HSSM layer with encrypted projections and full task model execution. It also required one explicit discardable HSSM warm-up row per tested horizon because the first HSSM row per horizon failed in the CKKS decode path. Measured rows succeeded after the warm-up workaround. The later multi-decay end-to-end runner extends this discipline by recording separate discardable warm-up rows for the single-decay and multi-decay HSSM paths.

The attention comparator is also deliberately bounded. The older softmax-free final-token power-attention row remains an operation-level feasibility check, and the cached polynomial-normalized rows report final-token and full-sequence variants with encrypted cached Q/K/V-style vectors. The stronger projected block adds homomorphic public-linear input and Q/K/V projections from encrypted bounded features at $T\in\{16,32\}$; these projections use public weights rather than model-private weights. The block still does not implement exact softmax, full transformer blocks, feed-forward layers, residual connections, layer normalization, model-private weights, or production inference. The current manuscript therefore compares against HE-friendly polynomial-attention primitives, not against full encrypted transformer systems such as THE-X, Powerformer, or MOAI.

For the labeled workflows, raw text is not stored. The old deterministic chunk-statistic feature tier is retained as calibration telemetry rather than task-quality evidence. The repaired tier uses frozen fastText embeddings on the client, train-normalized random projections, clipping, and public ridge readouts fit on training splits. Validation features are encrypted after projection and clipping; server-side HSSM or polynomial-attention computation is homomorphic; score decryption and thresholding are client-side. For the augmented-HSSM L40S rows, server encrypted HSSM evaluation accounts for 10.61--10.65\% of measured encryption/evaluation/decryption component time; this accounting excludes untimed plaintext tokenization, fastText lookup, projection, and clipping and does not support a stronger privacy claim.

\section{Experiments}

The experiments test whether the public-decay HSSM invariant changes FHE sequence-inference behavior under registered CPU, GPU, validation, and stress artifacts. The evidence is organized by claim type: audited deterministic features calibrate early telemetry, repaired fastText/L40S rows carry the task-quality result, and operation-level attention benchmarks isolate systems costs.

\subsection{Evidence Tiers}

Table~\ref{tab:evidence-tiers} summarizes the role of each experimental tier.

\begin{table*}[t]
\centering
\scriptsize
\caption{Evidence tiers and the supported interpretation for each experimental layer.}
\label{tab:evidence-tiers}
\setlength{\tabcolsep}{3pt}
\renewcommand{\arraystretch}{1.15}
\begin{tabularx}{\textwidth}{>{\hsize=0.55\hsize\linewidth=\hsize\raggedright\arraybackslash}X>{\hsize=0.55\hsize\linewidth=\hsize\raggedright\arraybackslash}X>{\hsize=1.90\hsize\linewidth=\hsize\raggedright\arraybackslash}X}
\toprule
Tier & Backend & Scope and supported reading \\
\midrule
Plaintext and CPU FHE checks & Python and OpenFHE CKKS & Structural sanity checks, packed-cell correctness, and a local public-carry timing advantage through $T=8$ with level-metadata caveats. \\

Audited toy-feature validation & FIDESlib CUDA & Calibration tier for latency/null-quality telemetry after the feature audit. \\

Repaired fastText HSSM validation & FIDESlib CUDA L40S & Encrypted HSSM execution from bounded projected fastText features through public readout, with exact plaintext/encrypted classification match. \\

Polynomial-attention comparators & FIDESlib CUDA L40S & HE-friendly cached final-token and full-sequence polynomial-normalized attention, plus a public-linear Q/K/V projection comparator; operation-level and task-readout scopes, not exact softmax or full transformer blocks. \\

Stress and variability rows & FIDESlib CUDA L40S & Context-length latency, materialized-score memory, and five-repeat timing stability for selected rows. \\

Causal-LM precursor & Plaintext quality plus encrypted audit files & Non-classifier plaintext quality comparison; encrypted candidate-set rows are raw audit only and are not claim-bearing LM quality, runtime, or memory metrics. \\
\bottomrule
\end{tabularx}
\end{table*}

\subsection{Operation Checks and Validation Repair}

The OpenFHE CPU single-cell benchmark successfully evaluates $h_{\mathrm{next}}=a h+g u$ on packed 8-slot encrypted vectors with maximum absolute error about $2\times10^{-12}$. The CPU recurrent-loop artifact then compares public-decay HSSM with naive encrypted selective recurrence over $T\in\{1,2,4,8\}$ under a depth-8 profile. At $T=8$, HSSM public-scalar carry operations sum to 23.710202 ms, while naive encrypted carry-gate-times-state operations sum to 165.264703 ms. The same artifact also records advancing OpenFHE level metadata, so this is a measured carry-operation timing result, not a backend-independent constant-level proof.

The initial full-validation tier used deterministic chunk-statistic features. Its audit shows why it cannot carry task-quality claims: toy-feature upper bounds remain near chance, while a simple plaintext lexical sanity check is much higher. Table~\ref{tab:validation-repair} records the decision.

\begin{table*}[t]
\centering
\scriptsize
\caption{Validation repair audit for the old deterministic feature tier. Toy-feature rows are retained as latency/null-quality telemetry only.}
\label{tab:validation-repair}
\setlength{\tabcolsep}{3pt}
\renewcommand{\arraystretch}{1.15}
\begin{tabularx}{\textwidth}{YYYYYY}
\toprule
Dataset & Majority & Best toy-feature validation & Toy-feature upper bound & Lexical sanity upper bound & Manuscript use \\
\midrule
Rotten Tomatoes & 0.5000 & 0.5563 & 0.5582 & 0.7739 & Demote task quality \\

SST-2 & 0.5092 & 0.5344 & 0.5436 & 0.8188 & Demote task quality \\
\bottomrule
\end{tabularx}
\end{table*}

The repaired task tier uses frozen fastText embeddings on the client, followed by train-normalized random projection, clipping, encryption, encrypted HSSM computation, homomorphic public readout, client decryption, and thresholding. The full encrypted HSSM forward-path audit maps this boundary explicitly: tokenization, embedding lookup, random projection, clipping, score decryption, and thresholding are client-side; bounded projected features, gate/write polynomials, local writes, public-decay aggregation, and public readout are covered by the encrypted server-side path. In the augmented-HSSM L40S task rows, each example enters the encrypted runner as $T=4$ chunks of width 128, or 512 bounded projected scalars; measured server-side encrypted HSSM evaluation accounts for 10.61--10.65\% of measured encryption, evaluation, and decryption component time. This is workload-partition evidence only, not a stronger privacy claim. The direct L40S HSSM validation reaches 0.750469 accuracy on Rotten Tomatoes and 0.741972 on SST-2, with 100\% plaintext/encrypted classification match, final score level 4 for every row, and no measured bootstrapping invocation.

\subsection{FastText Quality and Efficiency}

Table~\ref{tab:fasttext-joint} is the main non-toy task comparison. It compares an augmented public-decay HSSM readout with HE-friendly final-token and full-sequence polynomial-attention readouts under the repaired fastText workflow. All six encrypted rows have exact plaintext/encrypted classification match.

\begin{table*}[t]
\centering
\scriptsize
\caption{Repaired fastText L40S task quality and encrypted evaluation time. Final-token and full-sequence rows are polynomial-attention readouts, not exact softmax or full transformer blocks. Peak memory is descriptive row telemetry; logical state is the safe memory comparison within this task table.}
\label{tab:fasttext-joint}
\setlength{\tabcolsep}{3pt}
\renewcommand{\arraystretch}{1.15}
\begin{tabularx}{\textwidth}{YYYYYYY}
\toprule
Dataset & Model & Enc. accuracy & Class match & Mean eval & Peak memory & Logical state \\
\midrule
Rotten Tomatoes & HSSM & 0.7533 & 1.0000 & 15.22 ms & 1589 MiB & 5 \\

Rotten Tomatoes & Final-token attention & 0.7580 & 1.0000 & 11.81 ms & 1591 MiB & 9 \\

Rotten Tomatoes & Full-sequence attention & 0.7448 & 1.0000 & 75.24 ms & 2665 MiB & 12 \\

SST-2 & HSSM & 0.7511 & 1.0000 & 15.27 ms & 1589 MiB & 5 \\

SST-2 & Final-token attention & 0.7431 & 1.0000 & 11.86 ms & 1591 MiB & 9 \\

SST-2 & Full-sequence attention & 0.7397 & 1.0000 & 75.55 ms & 2665 MiB & 12 \\
\bottomrule
\end{tabularx}
\end{table*}

The peak-memory column in Table~\ref{tab:fasttext-joint} is retained as row-level process telemetry; the task-level memory comparison is the logical encrypted-state column. The table shows HSSM's main task-level advantage over the full-sequence polynomial-attention readout: HSSM is more accurate on both datasets, about $4.94$--$4.95\times$ faster, and uses a smaller logical state. It also identifies the final-token readout as a strong short-context baseline: that row is faster on these $T=4$ tasks and is 0.47 percentage points more accurate on Rotten Tomatoes, while HSSM is more accurate on SST-2.

\subsection{Polynomial-Attention Efficiency}

The first attention comparator is a polynomial-normalized attention primitive with encrypted cached Q/K/V-style vectors. It uses the Taylor kernel $1+z+0.5z^2$ and a low-degree reciprocal polynomial $1-r+r^2$ around the observed denominator mean. Table~\ref{tab:streaming-polynomial} reports the paired L40S operation-level comparison against the supported public-decay HSSM update.

\begin{table*}[t]
\centering
\scriptsize
\caption{Paired L40S operation-level comparison against HE-friendly polynomial attention. Physical peak-memory claims are not drawn from this table because the FIDESlib allocator pool keeps process peaks close.}
\label{tab:streaming-polynomial}
\setlength{\tabcolsep}{3pt}
\renewcommand{\arraystretch}{1.15}
\begin{tabularx}{\textwidth}{YYYYY}
\toprule
T & HSSM eval & Final-token attention eval & Full-sequence attention eval & Logical state: HSSM / final / full \\
\midrule
16 & 11.51 ms & 16.44 ms & 346.05 ms & 1 / 33 / 48 \\

32 & 22.48 ms & 31.25 ms & 1408.35 ms & 1 / 65 / 96 \\

64 & 44.91 ms & 60.02 ms & 5798.67 ms & 1 / 129 / 192 \\

128 & 91.84 ms & 148.88 ms & 23700.10 ms & 1 / 257 / 384 \\
\bottomrule
\end{tabularx}
\end{table*}

Across these paired rows, final-token polynomial attention is $1.34$--$1.62\times$ slower than HSSM, and full-sequence polynomial attention is $30.06$--$258.07\times$ slower. The same artifact supports lower logical encrypted-state footprint for HSSM. These cached-Q/K/V rows do not support projection-cost claims, exact softmax, layer normalization, feed-forward sublayers, or a full encrypted transformer block.

Table~\ref{tab:qkv-projection-comparator} reports the strengthened projection-cost comparator. Both rows succeed at $T=16$ and $T=32$ on L40S, but under different successful CKKS profiles. The comparison is operation-level only: the attention row covers encrypted public-linear input and Q/K/V projections, while no row measures exact softmax, a full transformer block, task accuracy, or model-private weights.

\begin{table*}[t]
\centering
\scriptsize
\caption{Compact projected-HSSM versus encrypted public-linear Q/K/V attention comparator on L40S. Mean evaluations are operation-level timings; exact softmax, full transformer-block behavior, task accuracy, and model-private weights are out of scope.}
\label{tab:qkv-projection-comparator}
\setlength{\tabcolsep}{3pt}
\renewcommand{\arraystretch}{1.15}
\begin{tabularx}{\textwidth}{YYYYY}
\toprule
Comparator & Successful T & Depth / ring & Mean eval, T=16 / T=32 & Peak memory \\
\midrule
Corrected projected HSSM & 16, 32 & depth 8, ring 32768 & 85.23 / 174.12 ms & 3628 MiB \\

Encrypted public-linear Q/K/V attention & 16, 32 & depth 10, ring 65536 & 156.90 / 380.62 ms & 19040 MiB \\
\bottomrule
\end{tabularx}
\end{table*}

Under these measured successful profiles, corrected projected HSSM has lower operation time and lower peak process memory than the encrypted public-linear Q/K/V attention block. The profile split is part of the systems result: projection cost is material, and the HSSM path reaches the measured lengths under the smaller depth-8/ring-32768 profile while the projected-attention path requires the larger successful profile in this artifact.

\subsection{Context-Length Stress and Variability}

The materialized quadratic-attention stress sweep measures context-length pressure rather than task quality. Under the fixed depth-8, scale-50, batch-8, ring-dimension-32768 profile, HSSM is evaluated in algebraically closed fixed-window form,

\[
h_T = \sum_t a^{T-1-t} g(x_t)u(x_t),
\]

which preserves the fixed-context HSSM state while avoiding repeated scalar multiplication of the carried ciphertext in FIDES/OpenFHE. The figure below shows the stress-sweep boundary: HSSM completes through $T=36$ on both GPUs; the RTX 4060 materialized quadratic-attention row fails at $T=36$; the L40S completes that row with a much larger physical memory footprint.

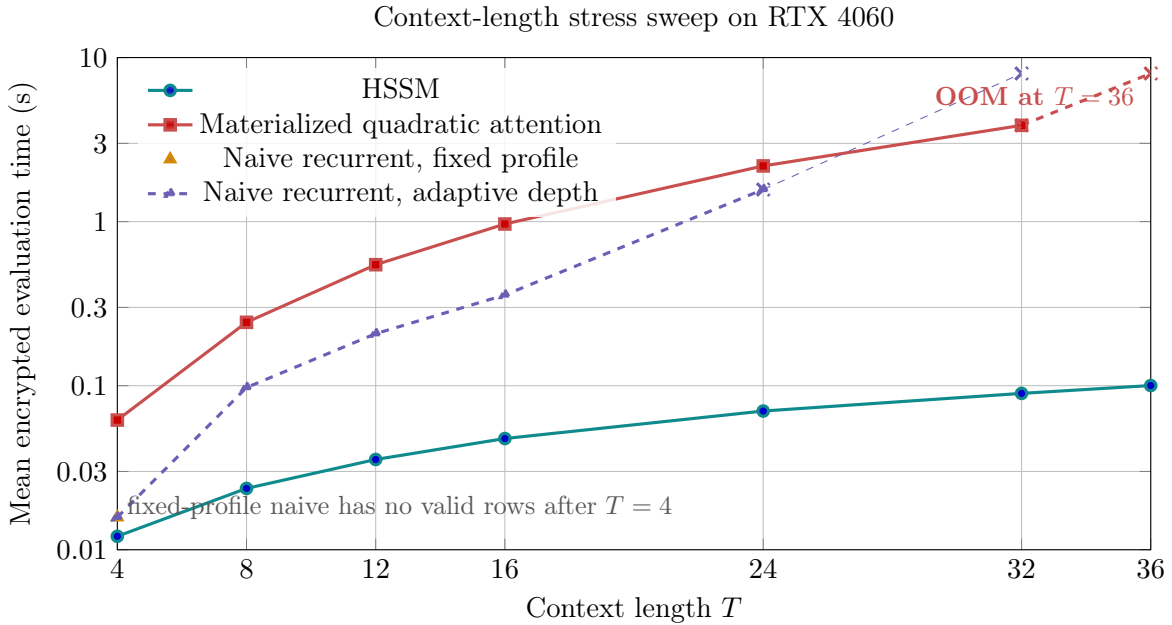
\begin{figure}[ht]
\centering
\begin{adjustbox}{max width=0.95\linewidth}
\begin{tikzpicture}
\definecolor{hssmteal}{HTML}{0F8B8D}
\definecolor{hssmamber}{HTML}{D58A00}
\definecolor{hssmred}{HTML}{C94F4F}
\definecolor{hssmviolet}{HTML}{6B5FB5}
\begin{semilogyaxis}[
  width=15.5cm,
  height=8.2cm,
  xlabel={Context length $T$},
  ylabel={Mean encrypted evaluation time (s)},
  xmin=4, xmax=36,
  ymin=0.01, ymax=10,
  xtick={4,8,12,16,24,32,36},
  ytick={0.01,0.03,0.1,0.3,1,3,10},
  log ticks with fixed point,
  grid=both,
  legend style={at={(0.02,0.98)}, anchor=north west, draw=none, fill=white, fill opacity=0.85, text opacity=1},
  title={Context-length stress sweep on RTX 4060},
]
\addplot+[hssmteal, very thick, mark=*] coordinates {
  (4,0.012079) (8,0.023648) (12,0.035501) (16,0.047625)
  (24,0.070063) (32,0.089754) (36,0.100207)
};
\addlegendentry{HSSM}
\addplot+[hssmred, very thick, mark=square*] coordinates {
  (4,0.061735) (8,0.243385) (12,0.546348) (16,0.967357)
  (24,2.185214) (32,3.863392)
};
\addlegendentry{Materialized quadratic attention}
\addplot+[hssmamber, very thick, only marks, mark=triangle*] coordinates {
  (4,0.015769)
};
\addlegendentry{Naive recurrent, fixed profile}
\addplot+[hssmviolet, very thick, dashed, mark=triangle*] coordinates {
  (4,0.015769) (8,0.097451) (12,0.207461) (16,0.357266) (24,1.569357)
};
\addlegendentry{Naive recurrent, adaptive depth}
\addplot+[hssmred, very thick, dashed, mark=x, mark options={scale=1.7, very thick}] coordinates {
  (32,3.863392) (36,8.0)
};
\node[anchor=north east, text=hssmred, font=\small\bfseries] at (axis cs:35.8,7.6)
  {OOM at $T=36$};
\addplot+[hssmviolet, dashed, mark=x, mark options={scale=1.7, very thick}] coordinates {
  (24,1.569357) (32,8.0)
};
\node[anchor=west, text=black!65, font=\small] at (axis cs:4,0.018)
  {fixed-profile naive has no valid rows after $T=4$};
\end{semilogyaxis}
\end{tikzpicture}
\end{adjustbox}
\caption{Context-length stress sweep under a fixed FIDESlib CUDA profile, with the adaptive-depth naive recurrent probe shown separately. HSSM completes through $T=36$, materialized quadratic attention reaches CUDA out-of-memory at $T=36$ on RTX 4060, and the L40S rerun completes the prepared materialized-attention row.}
\Description{Context-length stress sweep under a fixed FIDESlib CUDA profile, with the adaptive-depth naive recurrent probe shown separately. HSSM completes through $T=36$, materialized quadratic attention reaches CUDA out-of-memory at $T=36$ on RTX 4060, and the L40S rerun completes the prepared materialized-attention row.}
\end{figure}

Table~\ref{tab:l40s-stress} reports the L40S materialized stress rows. This is where physical GPU-memory separation is claimed; the paired polynomial-attention operation table above is used for latency and logical-state claims. The $T=4$ and $T=8$ stress rows sit near the allocator floor, so physical memory separation is read from the materialized rows where the quadratic score store becomes visible.

\begin{table*}[t]
\centering
\scriptsize
\caption{L40S stress sweep under the depth-8, scale-50, batch-8, ring-32768 profile. Materialized quadratic attention explicitly stores $T^2$ score ciphertexts.}
\label{tab:l40s-stress}
\setlength{\tabcolsep}{3pt}
\renewcommand{\arraystretch}{1.15}
\begin{tabularx}{\textwidth}{YYYYYY}
\toprule
T & HSSM mean eval & Quadratic mean eval & Quadratic / HSSM & HSSM peak & Quadratic peak \\
\midrule
4 & 5.74 ms & 24.14 ms & 4.20x & 1589 MiB & 1591 MiB \\

8 & 9.55 ms & 93.92 ms & 9.83x & 1589 MiB & 1591 MiB \\

12 & 12.62 ms & 212.79 ms & 16.86x & 1589 MiB & 2615 MiB \\

16 & 16.73 ms & 370.11 ms & 22.13x & 1589 MiB & 2615 MiB \\

24 & 23.54 ms & 846.51 ms & 35.95x & 1589 MiB & 4663 MiB \\

32 & 30.99 ms & 1455.83 ms & 46.98x & 1589 MiB & 7735 MiB \\

36 & 34.42 ms & 1884.77 ms & 54.77x & 1589 MiB & 9783 MiB \\
\bottomrule
\end{tabularx}
\end{table*}

Five independent L40S repeats support the stability of selected timing rows. At the direct $T=16$ overlap, final-token polynomial attention averages 16.47 ms with 0.15 ms 95\% half-width, full-sequence polynomial attention averages 341.84 ms with 4.78 ms half-width, and the HSSM stress row averages 16.75 ms with 0.11 ms half-width. At $T=36$, HSSM averages 34.89 ms with 0.12 ms half-width, while materialized quadratic attention averages 1881.96 ms with 12.21 ms half-width. These intervals apply only to the registered repeat rows, not to every timing in the paper.

\subsection{Multi-Decay and LM Boundaries}

The multi-decay decision is now resolved conservatively. In the old toy-feature tier, the two-decay setting [0.25, 0.75] on SST-2 improves over single-decay HSSM by 1.26 percentage points with 95\% interval [-0.23, 2.75], and the six-decay setting [0.1, 0.25, 0.5, 0.75, 0.9, 0.98] on Rotten Tomatoes improves by 0.28 percentage points with interval [-1.22, 1.78]. Those rows remain diagnostic/negative because the feature tier itself is weak. In the repaired fastText branch, the selected six-decay implementation reaches 0.7505 encrypted accuracy on Rotten Tomatoes and 0.7420 on SST-2 with exact encrypted/plaintext match; it supports the selected implementation, not broad multi-decay superiority.

The causal-language-model precursor is a forward-looking extension signal. In trained WikiText-2/BPE rows, the strongest recorded HSSM row reaches 17.78\% top-1 and 0.2670 MRR, while \texttt{poly\_attention\_causal\_full} reaches 12.76\% top-1 and 0.2247 MRR. The HE-ready plaintext branch also favors HSSM: 13.03\% top-1, 29.65\% top-5, and perplexity 183.9 versus 12.03\% top-1, 26.05\% top-5, and perplexity 206.0 for the polynomial-attention precursor. The encrypted candidate-set files are retained as audit material because a claim-bearing encrypted LM comparison still needs complete same-backend, same-candidate, same-profile rows.

\subsection{Experimental Reading}

The experimental reading is now strong and clean. The repaired fastText path gives non-toy encrypted L40S task utility with exact plaintext/encrypted classification match. HSSM has better task quality and about $5\times$ lower encrypted evaluation time than full-sequence polynomial attention in that workflow, and it has lower latency plus lower logical encrypted-state footprint than polynomial-attention variants in paired operation-level L40S rows. The projected-Q/K/V comparator adds operation-level evidence that projection cost is material under successful L40S profiles and that public-decay HSSM reaches the measured sequence lengths under a smaller successful CKKS profile. These results support the paper's target claim: public-decay carry paths are a measurable design lever for FHE-oriented sequence inference from encrypted bounded features. Exact softmax, full encrypted transformer blocks, model-private weights, encrypted raw-token-to-logit inference, measured bootstrapping schedules, and production private language-model serving remain outside that target claim.

\section{Discussion and Limitations}

The evidence supports HSSM as a focused FHE co-design result: keep the recurrent carry public/plaintext, confine encrypted multiplication to local write computation, and retain a fixed encrypted state rather than a context-growing attention cache or materialized attention matrix. The symbolic analysis separates HSSM from naive encrypted selective recurrence, the OpenFHE CPU loop shows a local timing advantage for public carry operations, the matched OpenFHE/FIDESlib trace aligns the same expanded HSSM path across backends, and the FIDESlib/L40S evidence now includes repaired fastText task rows plus stronger polynomial-attention comparators.

The central improvement over the initial evidence package is task quality. The old deterministic-feature validation tier is no longer treated as meaningful accuracy evidence. The repaired fastText branch gives non-toy encrypted L40S classification results with exact plaintext/encrypted match, and the joint fastText table shows HSSM beating full-sequence polynomial attention on both datasets while running about $5\times$ faster. Cached final-token polynomial attention remains a useful short-context baseline, especially on Rotten Tomatoes, but the central sequence-computation result is affirmative: when the comparison uses the full-sequence polynomial-attention readout, public-decay HSSM delivers higher task quality, lower encrypted evaluation time, and a smaller logical encrypted state.

The attention baseline is now stronger than the earlier softmax-free surrogate because the L40S comparator includes homomorphic public-linear input and encrypted Q/K/V projections from encrypted bounded features at $T \in \{16,32\}$. In that operation-level artifact, corrected projected HSSM succeeds under depth 8/ring dimension 32768, while the encrypted-Q/K/V final-token polynomial-attention block succeeds under depth 10/ring dimension 65536; under those measured profiles, HSSM is faster and uses lower peak process GPU memory. The comparator scope is deliberately HE-friendly attention/readout computation: polynomial attention with public projections and encrypted bounded-feature inputs. This scope is enough to test the paper's core systems question without claiming exact softmax, residual paths, layer normalization, feed-forward layers, model-private weights, full-vocabulary language-model decoding, or a production transformer system. Full encrypted systems such as THE-X, Powerformer, EncryptedLLM, MOAI, Cachemir, AEGIS, and PriTran remain the related systems context rather than the measured baseline class.

The implementation and threat-model boundaries are also explicit. The supported server-side encrypted fastText path begins after client-side tokenization, fastText lookup, train-normalized random projection, clipping, and encryption. For the augmented-HSSM L40S rows, $T=4$ and width is 128, so 512 bounded projected scalars per example enter the encrypted runner. The covered encrypted subgraph is bounded projected features through gate/write polynomials, local writes, public-decay aggregation, and homomorphic public readout, followed by client-side score decryption and thresholding. The measured workload split is useful for accounting: server encrypted HSSM evaluation accounts for 10.61--10.65\% of measured encryption/evaluation/decryption component time in these rows. The privacy claim rests on encrypting the projected features before server-side evaluation, not on the public projection itself. The active privacy setting remains single-client semi-honest encrypted inference with public model parameters; homomorphic tokenization, homomorphic embedding lookup, model-private-weight inference, fully homomorphic raw-token-to-logit decoding, and hiding sequence length, operation shape, ciphertext counts, timing, memory use, backend failures, public parameters, or access-pattern metadata are outside the hidden surface.

The level and bootstrapping results strengthen the implementation story. On the matched expanded-polynomial HSSM path at $T=8$, OpenFHE CPU and FIDESlib CUDA/L40S both finish at final level 3 with noise-scale degree 2 under depth 8, scale 50, and ring dimension 32768. The older OpenFHE zero-remaining-level observation came from a different streaming recurrent schedule, so it no longer clouds the matched-path interpretation. The fastText L40S HSSM rows complete under the depth-8 profile without invoking or measuring CKKS bootstrapping, and their final score telemetry reports level 4 for every row. That supports the measured no-bootstrap path for this bounded-feature HSSM configuration; bootstrap schedules, bootstrap latency, refresh policies, unbounded horizons, and larger HSSM variants are future scale-up questions.

The memory claim is positive but table-local. HSSM's fixed logical encrypted state is the safe comparison in the task and paired streaming tables. The task table's peak-memory column is descriptive row telemetry, and the paired streaming polynomial-attention table supports latency and logical-state claims rather than physical peak-memory superiority because allocator-pool behavior obscures process peaks. Physical peak-memory separation is supported only in the materialized-attention stress rows, with the short $T=4$ and $T=8$ cases treated as near allocator-floor measurements.

The causal-language-model precursor identifies a promising extension without distracting from the main contribution. The plaintext precursor shows that trained HSSM variants can beat a trained polynomial-attention precursor on WikiText-2/BPE ranking metrics, and the HE-ready plaintext rows preserve that ordering. A claim-bearing encrypted LM comparison should be a separate study: the current encrypted candidate-set files are useful audit artifacts, but the fixed-protocol comparison still needs complete same-backend, same-candidate, same-profile encrypted results or a full-vocabulary encrypted evaluation, plus a matched encrypted transformer-block baseline.

These boundaries define the next scale-up path: fuller encrypted transformer-block graphs, exact or better-calibrated softmax surrogates where feasible, measured bootstrapping schedules, fixed-protocol encrypted LM evaluation, longer contexts, and multi-GPU throughput. The current result is affirmative: in the L40S evidence, public-decay HSSM provides non-toy encrypted task utility, lower latency than full-sequence HE-friendly attention, lower logical encrypted-state growth, and a strengthened encrypted-Q/K/V comparator story under a clear and reproducible client/server boundary.

\section{Conclusion}

This paper establishes public-decay HSSM as a concrete FHE co-design pattern for server-side encrypted sequence computation from bounded features. Its central design move replaces the encrypted carry gate times encrypted carried state of a naive selective recurrence with public/plaintext recurrent decay, so ciphertext-ciphertext multiplication stays on the local write path instead of the recurrent carry path. The result is a fixed encrypted state across the sequence and a practical way to reduce encrypted carry-path pressure.

The evidence package gives this architecture a strong empirical base. After the deterministic-feature audit separates early telemetry from task-quality evidence, the repaired fastText/L40S workflow shows exact plaintext/encrypted HSSM classification match and non-toy task utility on Rotten Tomatoes and SST-2. In the same bounded-feature task setting, HSSM beats the full-sequence HE-friendly polynomial-attention comparator on both task rows while running about $5\times$ faster. The paired L40S operation-level rows add the systems signal: over the measured context lengths, public-decay HSSM has lower latency and lower logical encrypted-state footprint than both cached final-token and full-sequence polynomial attention.

The strengthened comparator and level evidence sharpen the result. The encrypted public-linear input and Q/K/V projection benchmark closes the projection-cost gap in the attention baseline: corrected projected HSSM succeeds at $T\in\{16,32\}$ under depth 8/ring 32768, while the encrypted-Q/K/V attention block succeeds under depth 10/ring 65536, with HSSM faster and using lower peak process GPU memory in the measured rows. The unified OpenFHE/FIDESlib trace shows that the same expanded HSSM path finishes at final level 3 and noise-scale degree 2 on both backends, so the level story is a matched-circuit result rather than an artifact of comparing different schedules.

The paper's explicit interface is a source of strength. In the task workflow, the client performs tokenization, frozen fastText lookup, projection, clipping, encryption, decryption, and thresholding; the server receives encrypted bounded projected features and evaluates gate/write polynomials, local writes, public-decay aggregation, and public readout. Within that interface, public-decay HSSM delivers non-toy encrypted task utility, fixed-state encrypted sequence computation, and measured latency/logical-state advantages against HE-friendly attention/readout comparators. The central contribution is therefore affirmative and measurable: public-decay carry paths are a practical design lever for FHE-oriented sequence inference.

\section*{Use of Generative AI Tools}

OpenAI language-model tools were used as editorial and coding assistants during manuscript preparation and workflow support. The authors take responsibility for all claims, experiments, citations, and text in this article.

\end{document}